\documentclass{article}
\usepackage[utf8]{inputenc}
\usepackage{xcolor}
\usepackage{amsmath}
\usepackage{amssymb}
\usepackage{authblk} 
\usepackage[left]{lineno}
\usepackage{lineno}
\usepackage{caption}
\usepackage{subcaption}
\usepackage[numbers,sort&compress]{natbib}
\usepackage{graphicx}
\usepackage{doi}
\DeclareUnicodeCharacter{2061}{}
\DeclareUnicodeCharacter{2032}{}
\DeclareUnicodeCharacter{2297}{}
\usepackage{enumerate}
\DeclareUnicodeCharacter{2060}{\nolinebreak}

\usepackage[utf8]{inputenc} 
\usepackage[T1]{fontenc}    
\usepackage{hyperref}       
\hypersetup{                              
    colorlinks=true,                
    breaklinks=true,                
    urlcolor= blue,                
    linkcolor= blue,                          
    bookmarksopen=false,
    filecolor=black,
    citecolor=blue,
    linkbordercolor=red}
\AtBeginDocument{\hypersetup{pdfborder={0 0 0}}}   
\usepackage{setspace}
\usepackage{geometry}
\title{\textbf{Journey through Gauge Theories: From Foundations to New Frontiers}}
\author[$1$]{Lisha} 
\author[$2$]{Suyash Mehta}
\author[$3$]{Neelu Mahajan\thanks{corresponding author: ibm.lhcms@gmail.com}}

\affil[$1$]{\textit{\small{Department of Physics, Goswami Ganesh Dutta College Sector-32C, Chandigarh, India}}}
\affil[$2$]{\textit{\small{Indian
Intitute of Space Science and Technology Valiamala Thiruvananthapuram, Kerala, India}}}
\affil[$3$]{\textit{\small{Department of Physics, Goswami Ganesh Dutta College Sector-32C, Chandigarh, India}}}
\date{}
\begin{document}
\maketitle
\begin{abstract}
From the humble beginnings of particle physics to groundbreaking discoveries, the journey of particle physics has revolutionized our understanding of the universe. The progression of particle physics, begin from the discovery of new particles, to gauge symmetries 
   and many successful theories in its journey which help in building up the model that explain the underlying makeup of matter, regulated by fundamental forces led to establishment of Standard Model. 
   The discovery of neutrino oscillations 
     compelled scientists to explore theories that extend beyond the Standard Model, which opened up new frontiers in the field of particle physics. The theories have shifted its paradigm to a single comprehensive theory at the scale of $10^{14}$ GeV. 
    This article explains the journey of particle physics from the basic constituent to the ``theory of everything.'' 

\end{abstract}
\section{Introduction}  Symmetry in particle physics are the clues that help us unravel the hidden patterns and secrets behind how everything works. From the smooth movements of electrons to the strong connections between quarks, studying these symmetries gives us a special way to uncover the secrets of the universe.
 The journey is a big step in understanding the electroweak model, which was an attempt to unify the electromagnetic and the weak nuclear force. As soon as the phenomena of spontaneous symmetry arise it becomes evident that symmetry is not the final answer to all the questions, giving rise to Standard Model(SM). Even after the tremendous success of SM, it falls short when it comes to unify all the forces, thus leading to the ``theory of everything''.

 In 1918, Hermann Weyl attempted to develop a unified theory that extends the geometric principles underlying General Relativity to encompass electromagnetism and gravitation, the theory in which the concept of linking the conservation of electric charge with gauge symmetry first emerged\cite{weyl1918gravitation}.
In 1929, Weyl reintroduced the gauge concept within the framework of quantum theory, introducing his ``gauge principle,'' which has exerted significant influence during the latter part of the century\cite{weyl1929electron}. It is well known that his attempt for unification failed\cite{brading2002symmetry} but Weyl published two significant papers in 1919\cite{weyl1919space} and 1929\cite{scholz2005local} both of which include his two-component theory of massless spin $\frac{1}{2}$ fermions\cite{straumann1996early}. The idea was abandoned after Einstein pointed out significant problems with the theory and it didn't reemerge until 1927 when Schrodinger introduced his wave equation for quantum theory in 1926\cite{Fock1926242} and complex wavefunctions had come to be recognized as having applications in physics\cite{aspect2017birth}. 
In 1934, inspired by Pauli's theory of beta decay in electromagnetism, Enrico Fermi proposed the first theory of beta decay, featuring a point interaction without reference to a force carrier.
Late in 1953, Yukawa on theoretical grounds proposed that every particle interaction requires a corresponding force field with the exchange of quantized particles mediating the interaction. Their existence was finally established by physicists in 1983\cite{vonpos}.
 
 In 1954, Robert Mills and Chen-Ning Yang introduced a non-abelian gauge field theory utilizing the isospin group, which led to their Nobel Prize-winning work in 1957\cite{poo2020conversation}. 
Yang and Mills introduced non-Abelian gauge theories in order to describe interactions between elementary particles using symmetries described by non-commuting mathematical groups and their work served as the cornerstone for the advancement of quantum chromodynamics (QCD)\cite{brink2015conference}. 

In 1961, it was discovered that hadrons could be arranged into triangularly symmetric, two-dimensional patterns of octets and decuplets. Gell-Mann and Yuval Ne'eman, Salam's students at the time, independently proposed an explanation for this by using a broader and more approximate $SU(3)$ symmetry which he called the ``eightfold way''\cite{gell2018eightfold}. The three lightest quarks u, d, and s were recognized as having this symmetry. 
In 1964, Englert Brout, Higgs and Guralnik proposed the same model for spontaneous symmetry breaking in the most basic $U(1)$ gauge theory which is a misconfigured form of electrodynamics of spin-0 charged particles\cite{guralnik1964global,higgs1964broken}. It entails the creation of a brand-new complex scalar field with the significant discovery of the Higgs boson and the mechanism that gave gauge boson masses while dodging the Goldstone theorem\cite{PhysRevD.98.045001} led to the final component of the puzzle in creating the SM 
(aside from the top quark, which completes the family of six).
Major experimental verification of SM was given by Large Hadron Collider (LHC) at CERN and the two devoted teams that designed, built and operated the magnificent ATLAS and CMS detectors whose main goal was to discover the Higgs boson. 
The search was triumphantly successful in 2012 and Englert and Higgs were awarded Nobel Prize in 2013 that clears the picture\cite{kibble2015spontaneous}.

Ever since its emergence, the SM has startling success to its credit. The SM of particle physics is a complete theory describing all the forces and the particles that interact through these forces. At its core, the SM is based upon Quantum Field Theory  thus, treating the particles as the excitations underlying the quantum fields. Despite of its impressive performance, there are several phenomenons that could not be explained by SM of particle physics. 
It does not incorporate gravity and also fails to provide an explanation for the existence of dark matter\cite{smith1990dark} which constitutes a significant portion of the universe's mass. Further, it does not offer a satisfactory explanation for the noted dissimilarity between matter and antimatter in the cosmos and many more. Physicists are actively working on the theories that go beyond the SM and addresses some of the unanswered questions. 

The article is mainly divided into five sections. Section 2 discusses about the Gauge theories and its applications on U(1), SU(2) and SU(3) symmetry, electroweak unification i.e. the  Glashow-Weinberg-Salam theory, spontaneous symmetry breaking that disrupted the uniformity of the universe and thus provided mass to the particles. Section 3 describes the SM of particle physics  that explains the interactions between the various fundamental particles and helps us to understand the intricate workings of the universe at even very small scale of about 100 GeV. Section 4 focuses on the  
theories that go beyond the SM, 
incorporating the new particles and discusses about the principles or forces to answer the unresolved questions of particle physics. Finally, Section 5 explains a single coherent framework that offers a comprehensive understanding of all the fundamental forces in nature.

\section{Gauge Theory}
Years passed, before physicists realized that gauge theories can adequately capture all of the known fundamental interactions. 
After 1958, there was a gradual realization that Yang-Mills gauge theory has the potential to effectively account for both the weak and strong interactions
\cite{RevModPhys.72.1}. Gauge theories, characterized by gauge symmetry, play a pivotal role in various quantum gravity theories and fundamental interactions, including Maxwell's electrodynamics. They maintain an unaltered Lagrangian under local gauge transformations and govern known interactions in nature. 
In view of a field $\Psi(x^u)$, the action integral is given as
\begin{equation}
   \mathcal{S}=\int\mathcal{L}(\Psi,\partial_u\Psi)d^4x,
\end{equation}
In physics philosophy, one's focus lies on internal symmetry transformations that preserve the distinct space-time properties of fields, ensuring that these transformations do not mix with the space-time components of the wave function. Taking $\Psi(x)\longrightarrow \exp(i\alpha)\Psi(x)$,
the family of phase transformation $\mathbf{U(x)}=\exp(i\alpha)$, where a single parameter $\alpha$ may continuously extend over real numbers,  forms a unitary Abelian group known as  $\mathbf{U(1)}$ group through Noether's theorem\cite{brading2000noether}, which implies the presence of a conserved current.
It is adequate to examine the invariance of $\mathcal{L}$ when subjected to an infinitesimal $\mathbf{U(1)}$ transformation. $\Psi\longrightarrow(1+i\alpha)\Psi$
so that $\mathcal{L}$ is unchanged, this is termed global gauge invariance. Incorporating the local gauge symmetry principle allows for interaction terms in the Lagrangian that multiply the fields when they interact, enabling a comprehensive description of experimental observations, with the parameter $\alpha$ varying with space-time dependence. 
The dependence of parameter $\alpha$ on `\textit{x}' is associated with space-time dependence with
$\Psi(x)\longrightarrow\exp(\alpha(ix))\Psi(x)$,
so the term $\partial_u\alpha$ breaks the local gauge invariance.

\subsection[Weyl's theory U(1) Symmetry]{Weyl's theory $\mathbf{U(1)}$ Symmetry}

In 1929, Weyl's papers were a turning point in the history. They articulated the contemporary concept of gauge invariance, asserting that the existence of the four vector potential is a consequence of the requirement for the matter equations to remain unchanged, when subjected to gauge transformations of the matter fields\cite{jackson2001historical}. 
QED is predicated on the idea that complex interactions involve numerous electrons and photons. 
Looking into the mathematical formalism, one begins with the free matter Dirac Lagrangian with a global gauge symmetry which is given as,
\begin{equation}
    \mathcal{L}=\hbar c\bar\Psi\gamma^u\partial_u\Psi+mc^2\bar\Psi\Psi;\hspace{1cm} where \hspace{1cm}\bar\Psi=i\Psi^\dagger\gamma^0.
\end{equation}
The Lagrangian remains invariant
under global gauge transformation but the same is not invariant under local gauge transformation.
 So one needs to replace $\partial_u\rightarrow\mathcal{D}_u=\partial_u+iq\mathcal{A}_u$ where $\mathcal{A}_u$ is a vector field. The term $\mathcal{A}_u$ transforms as $\mathcal{A}_u\rightarrow\mathcal{A}{'}_u$, hence the new Lagrangian can be written in terms of newly defined derivative as 
\begin{equation}
    \mathcal{L}{'}=\hbar c\bar\Psi\gamma^u\partial_u\Psi+i\hbar c\bar\Psi\gamma^u(\partial_u\phi(x^u))\Psi+iq\hbar c\bar\Psi\gamma^u\mathcal{A}{'}_u\Psi+mc^2\bar\Psi\Psi.
\end{equation} 
The interaction term between the fields comes out to be $iq\hbar\bar\Psi\gamma^u\mathcal{A}_u\Psi$. The new field $\mathcal{A}_u$ can't propagate, therefore to provide it a kinetic term, the Proca Lagrangian is written as
\begin{equation}
    \mathcal{L}_\mathrm{Proca}=\frac{1}{16\pi}\mathcal{F}_{uv}\mathcal{F}^{uv}+\frac{1}{8\pi}(\frac{mc}{\hbar})^2\mathcal{A}_{u}\mathcal{A}^{u}.
\end{equation}
 In order to impose local gauge symmetry, the Lagrangian needs to be invariant which is possible only if the mass term is zero then, the final Lagrangian becomes 
\begin{equation}
    \mathcal{L}=\hbar c\bar\Psi\gamma^u\partial_u\Psi+iq\hbar c\bar\Psi\gamma^u\mathcal{A}_u\Psi+mc^2\bar\Psi\Psi+\frac{1}{16\pi}\mathcal{F}_{uv}\mathcal{F}^{uv},
\end{equation}
where, q represents the charge of $\Psi$ field. The Lagrangian described above remains unchanged when subjected to
$\mathbf{U(1)}$ local gauge transformation and defines the electromagnetic four potential as well as the photon being the gauge boson 
\cite{zizzi20112}.

\subsection{Yang-Mills theory}

In 1954, Yang and Mills' papers, introduced non-Abelian gauge theories that provided a masterful overview of the origins and progression of gauge invariance\cite{gross1992gauge}. 
To extend the theory of the massless spin-1 field, which describes pure electromagnetism in a free field context, interactions through the application of Yang-Mills theory were introduced that allowed us to construct a framework for spin-1 fields that incorporates the dynamics of interactions.
The two spin-$\frac{1}{2}$ fields $\Psi_1$ and $\Psi_2$, for which the Lagrangian in the absence of interaction can be expressed as the sum of the two Dirac Lagrangian's 
\begin{equation}
    \mathcal{L}=[i\hbar c\bar\Psi_1\gamma^u\partial_u\Psi_1-m_1c^2\bar\Psi_1\Psi_1]+ [i\hbar c\bar\Psi_2\gamma^u\partial_u\Psi_2-m_2c^2\bar\Psi_2\Psi_2].
\end{equation}
When $\Psi_{1}$ and $\Psi_{2}$ were combined as column matrices and their adjoints as row matrices, the resulting Lagrangian takes on a new form as
\begin{equation}
    \mathcal{L}=i\hbar c\bar\Psi\gamma^u\partial_u\Psi-Mc^2\bar\Psi\Psi,
\end{equation}
where \textit{M} is $2\times2$ square matrix with $m_1$ and $m_2$ as principal diagonal elements.  
Consider parameter `a' be the function of $x^u$ then $a(x^u)=\frac{-\lambda(x^u) q}{\hbar c}$, where q is the coupling constant, the transformation $\Psi\rightarrow\Psi^{'}=e^\frac{-iq\sigma\lambda(x^u)}{\hbar c}\Psi\approx S \Psi$ is the local $\mathbf{SU(2)}$ transformation. Here, the Lagrangian is not invariant under such transformation as $\partial_u\Psi\rightarrow S\partial_u\Psi+(\partial_u S)\Psi$. Replacing $\partial_u$ with covariant derivative $\mathcal{D}_u=\partial_u+\frac{iq}{\hbar c}\sigma\mathcal{A}_u$, 
the Lagrangian becomes: 
\begin{equation}
     \mathcal{L}=[i\hbar c\bar\Psi\gamma^u\mathcal{D}_u\Psi-mc^2\bar\Psi\Psi]-(q\bar\Psi\gamma^u\sigma\Psi)\mathcal{A}_u.
\end{equation}
Hence, an interaction term ``$-(q\bar\Psi\gamma^u\sigma\Psi)\mathcal{A}_u$'' gets added to the Lagrangian  as $\mathbf{SU(2)}$ has 3 free parameters $\mathcal{A}^u=(\mathcal{A}^u_1,\mathcal{A}^u_2,\mathcal{A}^u_3)$, for propagation of new fields. When the Proca Lagrangian is taken into account, 
the Lagrangian becomes
\begin{equation}
    \mathcal{L}=[i\hbar c\bar\Psi\gamma^u\mathcal{D}_u\Psi-mc^2\bar\Psi\Psi]-\frac{1}{16\pi}\mathcal{F}^{uv}\mathcal{F}_{uv}-(q\bar\Psi\gamma^u\sigma\Psi)\mathcal{A}_u,
\end{equation}
which remained unchanged under local $\mathbf{SU(2)}$ gauge transformations and represents the interaction of two identical mass Dirac fields with three massless vector gauge fields. 

Considering the non-abelian $\mathbf{SU(3)}$ group, which exhibits symmetry when acted upon three-dimensional space through eight generators. The spinor field $\Psi$ is defined by three components $\Psi_r, \Psi_b, \Psi_g$ in color space, 
red, blue and green as the basis in three dimensions.
The Lagrangian of free Dirac field is given as: 
\begin{equation}
    \mathcal{L}=\hbar c\bar\Psi\gamma^u\partial_u\Psi+mc^2\bar\Psi\Psi\hspace{1cm} where \hspace{1cm} \bar\Psi=i\Psi^\dagger\gamma^0,
\end{equation}
the corresponding $\Psi$ is invariant globally under the following transformation
$\Psi \rightarrow \Psi' = e^{-i\frac{q}{\hbar c}\lambda \phi} \Psi$
where, $\lambda$ and $\phi$ correspond to matrix generator parameters respectively, having eight elements and $\frac{q}{\hbar c}=g$, which is a constant that determines the coupling strength of three types of color.
To ensure that the Lagrangian is locally invariant under $\mathbf{SU(3)}$, $\Psi\rightarrow\Psi^{'}=e^{-i\frac{q}{\hbar c}\lambda\phi(x^u)}\Psi$ the transformation $\mathcal{D}_u\Psi\rightarrow\mathcal{D}^{'}_u\Psi=e^{-ig\lambda\phi}$ were performed and a term $ig\hbar c\Bar{\Psi}\gamma^u\lambda\mathcal{A}_u\Psi$ got added to the Lagrangian after transformation.
For propagation of $A_{u}$, the massless Proca Lagrangian was taken into account. The massless Proca Lagrangian $\frac{1}{16\pi}\mathcal{F}_{uv}\mathcal{F}^{uv}$ with $\mathcal{F}_{uv}=-\frac{i}{g}[\mathcal{D}_u,\mathcal{D}_v]$ was used so as to keep the Lagrangian invariant. Solving the final value of the kinetic term for eight fields $\mathcal{A}^a_u$, resulted in,
\begin{equation}
    \frac{1}{16\pi}\mathcal{F}_{uv}\mathcal{F}^{uv}=\frac{1}{16\pi}(\partial_u\mathcal{A}^a_v-\partial_v\mathcal{A}^a_u)(\partial^u\mathcal{A}^{av}-\partial^v\mathcal{A}^{au})+ \textit{gluon-gluon interactions},
\end{equation}
here $a\hspace{0.1cm}\in\hspace{0.1cm} \{{1,.......,8}\}$, the gluon-gluon interactions were due to the fact that $\mathbf{SU(3)}$ was non-abelian such that structure constant $f^{abc}\neq0$. 
Hence, the final Lagrangian is given as:
\begin{equation}
\begin{aligned}
    \mathcal{L} &=\hbar c\bar\Psi\gamma^u\partial_u\Psi+mc^2\bar\Psi\Psi- \frac{g}{16\pi}f^{ade}\mathcal{A}^d_u\mathcal{A}^e_v(\partial^u\mathcal{A}^{va}-\partial^v\mathcal{A}^{ua})\\ 
    & -\frac{g}{16\pi}f^{abc}\mathcal{A}^{ud}\mathcal{A}^{vc}(\partial_u\mathcal{A}^a_v-\partial^v\mathcal{A}^a_u)\\ 
    & +\frac{g^2}{16\pi}f^{abc}f^{ade}\mathcal{A}^a_u\mathcal{A}^c_v\mathcal{A}^{ud}\mathcal{A}^{ve}. 
\end{aligned}
\end{equation}

In summary, the fundamental difference between abelian and non-abelian gauge symmetries is rooted in the nature of gauge transformations. Abelian symmetries exhibit commutative transformations, while non-abelian symmetries lack this property.
By understanding these gauge symmetries one can pave a deep understanding of electroweak model, an attempt to unify these abelian and non-abelian gauge symmetries.

\subsection{Electroweak unification}
In 1961, Sheldon Glashow introduced $SU(2)\otimes U(1)$ gauge theoretical electroweak interaction model. The model is based on a simple application of the gauge principle: the notion that an inhomogeneous coupling term must exist to fulfill the condition of the fundamental Lagrangian's invariance under local gauge transformations\cite{lyre2008does}.
The theory is developed long before such exact measurements are possible, 
the electroweak unification (GWS theory) accurately describes electroweak interactions up to the currently tested energies\cite{aitchison2012gauge}.
Beginning with the free Dirac Lagrangian  
and assuming $\Psi$ as Weyl spinor  $\Psi_{\pm}$,  the Dirac Lagrangian is written as,
\begin{equation}
    \mathcal{L}=-\hbar c(i\Psi^\dagger_{-}\partial_u\sigma^u\Psi_{-}+i\Psi^\dagger_{+}\partial_u\bar\sigma^u\Psi_{+})+ mc^2(\Psi^\dagger_{-}\Psi_{+}+\Psi^\dagger_{+}\Psi_{-}),
\end{equation}
where, $\sigma^u=(I,\sigma^i)$ and $\bar\sigma^u=(I,-\sigma^i)$ and the four component $\Psi$,
the final Lagrangian becomes
\begin{equation}
    \mathcal{L}=\hbar c\bar\Psi_L\gamma^u\partial_u\Psi_L+\hbar c\bar\Psi_R\gamma^u\partial_u\Psi_R+mc^2(\bar\Psi_L\Psi_R+\bar\Psi_R\Psi_L).
\end{equation}
Considering the first generation of leptons and omitting the mass term for the sake of simplicity, the above Lagrangian could be written as
\begin{equation}
     \mathcal{L}=\hbar c\bar\chi_L\gamma^u\partial_u\chi_L+\hbar c\bar e_R\gamma^u\partial_u e_R.
\end{equation}
 The Lagrangian remains invariant in global symmetry of $\mathbf{SU(2)_L\otimes U(1)_Y}$ as $\mathbf{SU(2)_L}:\chi_L\rightarrow\chi^{'}_L=e^{-ig\sigma\theta}\chi_L$ and $\mathbf{U(1)_Y:}\chi_L\rightarrow\chi^{'}_L=e^{-ig^{'}Y_{\chi_L}\phi}\chi_L,\hspace{0.1cm}e_R\rightarrow e^{'}_R=e^{-ig^{'}Y_{e_R}\phi}$ where, $\sigma$ is the generator and $\theta$ is parameter, $g$ denotes the coupling.
Applying local symmetry transformation to Lagrangian, the quantities $\theta$ and $\phi$ depend on $x^u$, so
$\partial_u\chi_L\rightarrow\mathcal{D}_u\chi_L = \partial_u\chi_L+ig\sigma\mathcal{W}_u\chi_L+ig^{'}\mathcal{Y}_{\chi_L}\mathcal{B}_u\chi_L$ and $\partial_u e_R\rightarrow\mathcal{D}_u e_R = \partial_u e_R+ig^{'}\mathcal{Y}_{e_R}\mathcal{B}_u e_R$. Here, $\mathcal{W}_u$ are three gauge field of $\mathbf{SU(2)_L}$ and $\mathcal{B}_u$ is for $\mathbf{U(1)_Y}$ with one gauge field. The transformation of fields is given by
$\sigma\mathcal{W}_u\rightarrow\sigma\mathcal{W}^{'}_u=e^{-ig\sigma\theta}\sigma\mathcal{W}_u e^{ig\sigma\theta}+\frac{i}{g}\partial_u(e^{-ig\sigma\theta})e^{ig\sigma\theta},\hspace{0.1cm} \mathcal{B}_u\rightarrow\mathcal{B}^{'}_u=\mathcal{B}_u+\partial_u\phi$
and the covariant derivative works perfectly under such a transformation. $\mathcal{F}_{uv}=-\frac{i}{g}[\mathcal{D}_u,\mathcal{D}_v]$, term is being used to add a kinetic term for each gauge field as
\begin{equation}
    \mathcal{F}_{uv}=\partial_u\mathcal{B}_v-\partial_v\mathcal{B}_u,\hspace{1cm}\mathcal{F}^a_{uv}=\partial_u\mathcal{W}^a_v-\partial_v\mathcal{W}^a_u-g\epsilon^{abc}\mathcal{W}^b_u\mathcal{W}^c_v,
\end{equation}
where, $\epsilon^{abc}$ represent structure constant of  $\mathbf{SU(2)}$. Three gauge bosons $\mathcal{W}^1_u,\mathcal{W}^2_u,\mathcal{W}^3_u$ interact with the doublets and are defined as $\mathcal{W}^z_u=\mathcal{W}^3_u$ and $\mathcal{W}^\pm_u=\frac{1}{2}(\mathcal{W}^1_u\pm i\mathcal{W}^2_u)$\cite{Altarelli2020}.
In terms of the gauge principle,
only left chiral fermions are involved in the charged current interactions, which has a vector axial structure. Due to this electroweak symmetry breaking mechanism Z boson do not exhibit V-A couplings with fermions \cite{godbole2017field}. 
Still, this theory was unable to explain one of the major problems, the invariance of Proca mass term i.e. $\mathcal{W}^a_u\mathcal{W}^{ua}$ because the weak gauge bosons are massive and for providing mass term to the spinor, the left and right parts that belong to $\Psi$ needed to be combined but the gauge theory of electroweak interaction affected the left and right parts of the spinor differently.
These problems resulted in both gauge bosons and fermions being massless, which was contradictory to real values, the anomaly was further resolved by Peter Higg, Brout, Englert and Kibble in 1964 as they strechout the work by Goldstone on spontaneous symmetry breaking (SSB) to gauge theories\cite{lyre2008does}.

\subsection{Spontaneous symmetry breaking}
At an earlier time in the universe, everything is massless and this implies that 
$\mathbf{SU(2)_L}\otimes \mathbf{U(1)_Y}\rightarrow\mathbf {U(1)_{EM}}$ is the symmetry of the SM. As the universe underwent cooling, the symmetry of the universe breakdown to $\mathbf{U(1)_{EM}}$. While undergoing this process, three of the gauge bosons and the matter particles gained mass and these undesired massless scalar bosons, leading to Spontaneous symmetry breaking(SSB)\cite{kibble2015spontaneous}.
Considering $\mathbf{U(1)}$ symmetry, the free Lagrangian in terms of scalar field $\Phi=\Phi_1+i\Phi_2$ is given as 
\begin{equation}
    \mathcal{L}=\frac{1}{2}(\partial_u\Phi)^\ast(\partial^u\Phi)-\frac{1}{2}\mu^2\Phi^\ast\Phi+\frac{1}{4}\lambda^2(\Phi^\ast\Phi)^2,
\end{equation}
and is invariant under the transformation. Here, $-\frac{1}{2}\mu^2\Phi^\ast\Phi$ is the mass term\cite{christensen2009feynrules} and the term $\frac{1}{4}\lambda^2(\Phi^\ast\Phi)^2$ is the self interaction term. Localizing it as $\Phi(x^u)$ and making the covariant derivatives as $\mathcal{D}_u=\partial_u+i\frac{q}{\hbar c}\mathcal{A}_u$ and the gauge fields transforms as $\mathcal{A}_u\rightarrow\mathcal{A}^{'}_u=\mathcal{A}_u-\frac{q}{\hbar c}\partial_u\theta$. The final Lagrangian was written with the addition of kinetic term from the Proca Lagrangian,  
\begin{equation}
    \mathcal{L}=\frac{1}{2}[(\partial_u-\frac{iq}{\hbar c}\mathcal{A}_u)\Phi^\ast][(\partial^u+\frac{iq}{\hbar c}\mathcal{A}^u)\Phi)]-\frac{1}{2}\mu^2\Phi^\ast\Phi+\frac{1}{4}\lambda^2(\Phi^\ast\Phi)^2+\frac{1}{16\pi}\mathcal{F}^{uv}\mathcal{F}_{uv}.
\end{equation}
 The final Lagrangian remains invariant under the transformations.  
This theory introduces the concept of massless gauge fields
and the tiny fluctuations in the field defined the observable particles. 
The three fields with the tiny fluctuation is written as, $\mathcal{A}_u = 0+\delta\mathcal{A}_u, \Phi_1 = \frac{\mu}{\lambda}+\delta\Phi_1=\frac{\mu}{\lambda}+\eta, \Phi_2 = 0+\delta\Phi^\ast_2=\beta$ and on substituting these values in equation (2.18) Lagrangian  becomes
    \begin{equation}
     \mathcal{L}= [\frac{1}{2}(\partial_u\eta)(\partial^u\eta)+\mu^2\eta^2]+[\frac{1}{2}(\partial_u\beta)(\partial^u\beta)]+[\frac{1}{16\pi}\mathcal{F}^uv\mathcal{F}^uv+\frac{1}{2}(\frac{q\mu}{\hbar c\lambda})^2\mathcal{A}_u\mathcal{A}^u]
        +other.
    \end{equation}
Hence, one has a massive real scalar field $\eta$ with $m_\eta=\frac{\sqrt{2}\hbar}{c}\mu$ called Higgs field and a massive gauge field $\mathcal{A}_u$ with $m_A=2\sqrt{\pi}(\frac{q\mu}{\lambda c^2})$. In addition to these, physicists also got a massless scalar field $\beta$ called Goldstone boson\cite{beekman2019introduction}. 
Thus, the Higgs mechanism gave mass to the gauge fields by spontaneous symmetry breaking through coupling to an extra Higgs field $\Phi$. 
In addition, the electroweak theory of mass generation of the $W^\pm$ and $Z^0$ gauge bosons is made possible only by the Higgs mechanism and that too with significant physical impacts\cite{pittphilsci9295}.
It is easier to understand particle interactions when three forces 
(electromagnetic, weak and strong) are combined into a single framework, as is the case with the SM. This framework is supported by experimental data and has proven to be a potent tool for describing the behaviour of particles at high energies.

\section{The Standard Model}
The most successful theory in particle physics that describes all the fundamental phenomenon in nature and that has been verified experimentally to a high degree of precision is the Standard Model. It has revolutionized our thinking in understanding the behaviour of matter and energy at a sub-atomic level. The makeup of SM began in 1970, when Sheldon L. Glashow, John Iliopoulos and Luciano Maiani proposed the GIM Mechanism\cite{maiani2013gim}, which replaces the $\delta S = 1$ neutral current terms with a new term resulting from the presence of a new quark(charm, ``c'').  It contributed as a crucial component in the unified theories of electromagnetic and weak interactions. Further, in 1974, the fundamental particle $J/\Psi$ that plays an important role in confirming the quark model of the particle was discovered by a group led by physicists Samuel Ting and Burton Richter, two separate research groups working at Stanford Linear Accelerator Center (SLAC) and the Brookhaven National Laboratory respectively\cite{dalitz1977introductory}.
In the late 20th century, particle physics achieved a significant milestone with the Standard Model (SM). The SM classifies elementary particles into two main categories: bosons with spin integral spin, responsible for mediating fundamental forces and fermions, which include quarks and leptons and constitute matter and posses half integral spin.
\cite{novaes2000standard,herrero1999standard,inbook,RevModPhys.71.S96}. 

The mathematical expression describing the above fundamental particle is a gauge theory based on group 
$\mathbf{SU(3)_c\otimes\mathbf{SU(2)_L}\otimes\mathbf{U(1)_Y}}$. The group $SU(3)_c$ where c stands for color degree of freedom, describes strong interaction of quraks and antiquarks mediated through octet of massless, intermediate vector bosons called gluons. The product $SU(2)_L \otimes U(1)_Y$ where, L and Y represents left handed and hypercharge respectively and describes the electroweak model that unifies electromagnetic and weak interaction mediated through massless photon and massive vector bosons $\mathcal{W}^\pm,\mathcal{Z}^0$. In the SM, quarks, charged leptons are massive particles while neutrinos are massless and exist only in left handed helicity state.
To construct the final Lagrangian for the SM, all the necessary components, including the terms describing electroweak, strong and Higgs interactions that give mass to particles through electroweak symmetry breaking, are combined. The Lagrangian describing the SM is then given as :

    \begin{align}
\begin{split}
\mathcal{L}_{SM} = &-\frac{1}{4}\mathcal{W}_{uv}\mathcal{W}^{uv} -\frac{1}{4}\mathcal{B}_{uv}\mathcal{B}^{uv}
+\bar L\gamma^u\left(i\partial_u-g\frac{1}{2}\sigma\mathcal{W}_u-g^{'}\frac{Y}{2}\mathcal{B}_u\right)L\\
&+\bar R\gamma^u\left(i\partial_u-g^{'}\frac{Y}{2}\mathcal{B}_u\right)R +\left|\left(i\partial_u-g\frac{1}{2}\sigma\mathcal{W}_u-g^{'}\frac{Y}{2}\mathcal{B}_u\right)\Phi\right|^2-V(\Phi)\\
&-(G_1 \bar L\Phi R+G_2\bar L\Phi_c R +\text{hermitian conjugate}).
\end{split}
\end{align}
The above Lagrangian incorporates kinetic energy, self-interaction as well as interactions with $\mathcal{W}^\pm,\mathcal{Z}^0,\gamma$ for leptons and quarks, including Higgs-related masses, couplings and connections to lepton and quark masses.
One way to categorize these independent parameters within the SM for a single generation is by classifying them as the inherent variables of the model as
the two gauge couplings $g,g^{'}$ for the $\mathbf{SU(2)_L}$ and $\mathbf{U(1)_Y}$ gauge groups respectively, the potential function $V(\Phi)$ involves two parameters namely $\mu$ and $\lambda$ and  
Y represents the Yukawa couplings.
The addition of two  generations of quarks and leptons, the SM consists of 25 free parameters including gravity and if gravitational forces are not taken into account it consists of 18 free parameters which include 3 coupling constants, 2 parameters in Higgs potential, 6 quark masses, 3 charged lepton masses, 3 mixing parameters, 1 CP-violating phase $\delta$. 
Major successes of the SM were the observation of neutral current interaction in 1983, the discovery of $\mathcal{W}^\pm$ and $\mathcal{Z}^0$ at CERN\cite{Watkins:1986va}, which were the crucial experiments that gave impetus to SM. 
The discovery of top quark by the CDF group
and DO group at FNAL\cite{campagnari1997discovery}, the crucial experiments that gave concreteness to the SM.
Thus, a long time was invested in developing the entire SM i.e. from the discovery of electron in 1897 by physicist J.J. Thomson, to the Higgs Boson, the last piece of the puzzle, which was discovered in 2012 at the Large Hadron Collider\cite{article}. Beside the remarkable achievement of SM of particle physics it failed to explain the observation of the neutrino oscillations by Super Kamiokande\cite{fukuda2003super} and SNO\cite{mcdonald2021neutrino}, which implies massive and non-degeneracy of neutrino. It does not address the presence of dark matter and dark energy which constitute approximately 95 $\%$ of the universe's mass and energy.
It does not include gravity and thus fails to provide a single unified theory that could explain all the fundamental forces in nature. Thus, to unravel physics beyond SM, vast and countless efforts have been put forward both on theoretical and experimental front in the hope of arriving at the so called ``theory of everything''.
\section{Beyond Standard Model}

 The SM of particle physics has been successful in explaining and estimating various natural phenomenon, but there are several reasons why we need to extend our theories Beyond Standard Model (BSM). This section, describes the theories which not only provide answers to unresolved problems in particle physics but also allow us to acquire a more profound understanding of the fundamental nature of the universe. By examining these theories, scientists can propose new experiments and reinterpret existing data to search for new particles and phenomena. This exploration also requires the development of advanced techniques and technologies including particle detectors and accelerators. By pushing the boundaries of our understanding, we can revolutionize physics, discover new avenues for research and establish connections between different branches of science. Thus, the need to go BSM is crucial for advancing our knowledge and exploring the mysteries of the universe.

\subsection{Neutrino Oscillations}

Neutrino oscillations refer to the phenomenon where neutrinos change their flavor as they travel, transitioning between electron, muon and tau flavors. The discovery provided compelling evidence indicating that neutrinos possess mass which necessitated an expansion of the SM. The observation of neutrino oscillations has a rich history over the decades. Several significant efforts were made both theoretically and experimentally in order to explain this phenomenon from the time it was observed in 1949 to the time it was discovered in 2000. In the early stages, while studying solar neutrinos, scientists noticed that the number of electron neutrinos reaching Earth were less than expected. The shortfall in the number of electron neutrinos was discovered by Raymond Davis Jr. at the Homestake Gold Mine\cite{ahluwalia1998reconciling}. This inconsistency was recognized as the ``solar neutrino problem.''

  Pontecorvo's Theory in 1957 suggested that neutrino exists in different flavor states and undergo a transition as they propagate through the vacuum or medium\cite{pontecorvo1958mesonium}. In 1962, Ziro Maki, Masami Nakagawa and Shoichi Sakata  established a relationship between the flavor eigenstates ($\upsilon_{e},\upsilon_{\mu},\upsilon_{\tau}$) and the mass eigenstates of neutrinos ($\upsilon_{1},\upsilon_{2},\upsilon_{3}$).  This neutrino mixing phenomenon is characterized by the Pontecorvo-Maki-Nakagawa-Sakata (PMNS) matrix\cite{rodejohann2015origin}.
\begin{equation}
\begin{pmatrix}
\upsilon_e \\
\upsilon_\mu \\
\upsilon_\tau
\end{pmatrix}
=
\begin{pmatrix}
U_{e1} & U_{e2} & U_{e3} \\
U_{\mu1} & U_{\mu2} & U_{\mu3} \\
U_{\tau1} & U_{\tau2} & U_{\tau3}
\end{pmatrix}
\begin{pmatrix}
\upsilon_1 \\
\upsilon_2 \\
\upsilon_3
\end{pmatrix},
\end{equation}

where, $U_{e1}$,  $U_{\mu1}$, $U_{\tau1}$ are the elements of the PMNS matrix which corresponds to mixing between the flavor and mass eigenstates. This oscillation concept can also be expressed as the probability of transition between different neutrino flavor states during their journey through space in terms of mixing angles ($\theta_{13}$, $\theta_{23}$) and the CP Violating phase ($\delta$).

\begin{align}
    P(\nu_\alpha \rightarrow \nu_{\beta})&= |U_{\beta1}|^2 \cdot |U_{\alpha1}|^2 \cdot \sin^2(\Delta m^2_{21} L / 4E) \notag \\&\quad + |U_{\beta2}|^2 \cdot |U_{\alpha2}|^2 \cdot \sin^2(\Delta m^2_{32} L / 4E) \notag \\
    &\quad + |U_{\beta3}|^2 \cdot |U_{\alpha3}|^2 \cdot \sin^2(\Delta m^2_{31} L / 4E).
\end{align}
where, the mixing between the $\alpha$-th flavor and the $i$-th mass eigenstate is denoted as $U_{\alpha i}$. The mass squared differences of the eigenstates are represented by $\Delta m_{21}^2$, $\Delta m_{32}^2$, and $\Delta m_{31}^2$. The variables \textit{L}
 and \textit{E} correspond to the distance travelled by neutrinos and the energy of neutrinos, respectively.

The first direct evidence of neutrino flavor change was observed in an experiment performed by Leon Lederman, Melvin Schwartz and Jack Steinberger \cite{steinberger1964leon} at the Brookhaven National Laboratory in the year 1962.  A deficit in the number of muon neutrinos was detected when these muon neutrinos were produced along with pions indicating the flavor change of muon neutrinos in the other flavor.
In 1973, experiments like The Gargamelle experiment at CERN, the IMB experiment and the Kamiokande experiment verified the shortfall of atmospheric muon neutrinos and offered a substantiating proof of neutrino oscillations\cite{hirata1990constraints,schlatter2015highlights}.  Physicists L. Wolfenstein, S. Mikheyev, and A. Smirnov introduced the Mikheyev-Smirnov-Wolfenstein (MSW) effect in 1978\cite{roulet1991mikheyev}. They showed that neutrinos propagating through matter, as the dense core of the sun can experience resonant flavor conversions due to interactions with electrons. The flavor oscillation probability when neutrinos propagate through matter is : 
\begin{align}
    P(\nu_e \rightarrow \nu_{\mu}) &= 1 - \sin^2(2\theta_{13}) \cdot \sin^2\left(\frac{{\Delta m^2_{31} \cdot L}}{{4E}}\right) \notag \\&\quad - \cos^4(\theta_{13}) \cdot \sin^2(2\theta_{12}) \cdot \sin^2\left(\frac{{\Delta m^2_{21} \cdot L}}{{4E}}\right).
\end{align}
In 1996 the experiments like Super-Kamiokande\cite{yasuda1996three} and SNO\cite{bandyopadhyay2003neutrino} were the pioneering experiments that provided precise results on the measurements of calculations of mixing angles. In 1998, the major goal of neutrino studies was the neutrino mass hierarchy whether normal $(m_{1}<m_{2}<m_{3})$ or inverted $(m_{3}<m_{1}<m_{2})$\cite{klapdor2001evidence}. C. Giunti, C.W. Kim and M. Monteno  presented a general scheme on three massive neutrinos and mass hierarchy with the comprehensive formalism describing the neutrino oscillations in earth\cite{giunti1998atmospheric}. Experiments like KamLAND and atmospheric neutrino experiments measured the mass squared differences and provided a hint towards the mass hierarchy\cite{li2013unambiguous}. Andre de Gouvêa, Boris Kayser and Stephen Parke provided the theoretical understanding of leptonic CP violation in neutrino oscillations and its connection to matter-antimatter asymmetry in the universe in 2003\cite{de2003manifest}.
Karsten Heeger, Jonathan Engel and Shun Zhou in 2005 introduced sterile neutrinos and explained their impact on neutrino oscillations \cite{de2005neutrino}. In 2012, physicists like Stephen Parke and Boris Kayser were pivotal in the refinement of the theoretical framework for neutrino oscillations. They used data from experiments like T2K and MINOS and established on theoretical basis the non-zero value of $\theta_{13}$ \cite{adler2012ghost}.

Year 2015 is a remarkable year when Takaaki Kajita and Arthur B. McDonald received the Nobel Prize in Physics for their revelation on neutrino oscillations\cite{ohlsson2016special}. Theoretical aspects of neutrino mass scale determination and the connection between cosmological observations and neutrino oscillation measurements were explained by  Kathrin Valerius, Shunsaku Horiuchi and Thomas Schwetz-Mangold in 2016\cite{roach2023long} whereas the theoretical understanding of neutrino mass ordering, non-standard neutrino interactions and the implications for future experimental measurements was explained by Alexander Friedland, Giancarlo Fiorentini and Walter Winter in 2022\cite{arrington2022physics,horiuchi2011cosmic}.

Thus, in order to explain the phenomenon of neutrino oscillations, scientists have developed various theoretical models including texture zeros, radiative neutrino mass models, the inverse seesaw mechanism, quasi-dirac neutrinos and flavor symmetry models\cite{nakamura2017towards,desai2003three,anamiati2019quasi}. These models aim to provide theoretical frameworks that can explain the detected patterns of neutrino oscillations and illuminate the fundamental mechanisms behind neutrino mass generation and flavor mixing. 

The neutrino mass term which was originally missing in the Lagrangian of SM was now included and the final Lagrangian is written as :
\begin{equation}
    \mathcal{L} = \mathcal{L}_{\text{SM}} + \sum_i m_i \bar{\psi}_i \psi_i - \sum_{\alpha,\beta} U_{\alpha i} U_{\beta i} \bar{\psi}_\alpha \gamma_\mu (1 - \gamma^5) \psi_\beta Z^\mu,
\end{equation}
where, $\mathcal{L}_{\text{SM}}$ represents the Lagrangian of the SM without the neutrino mass term.
The first term $\sum_i m_i \bar{\psi}_i \psi_i$ corresponds to mass term for neutrinos, where $m_{i}$ is the mass of the i-th neutrino flavor and $\psi_{i}$ represents the field linked with that flavor.
The second term $\sum_{\alpha,\beta} U_{\alpha i} U_{\beta i} \bar{\psi}_\alpha \gamma_\mu (1 - \gamma^5) \psi_\beta Z^\mu$ represents the mixing term, where $U_{\alpha i}$ is the element of the PMNS matrix corresponding to the mixing between the $\alpha$-th and i-th flavors, $\psi_{\alpha}$ \text{and} $\psi_{\beta}$ are the fields associated with the $\alpha$ -th and $\beta$-th neutrino flavors, $\gamma_\mu$ is the gamma matrix and $Z^\mu$ represents the $Z$ boson field.

While on the experimental side, a number of experiments including Daya Bay, CHOOZ, Kamland, RENO, MINOS and NOvA were conducted to study neutrino oscillations\cite{cao2016overview,apollonio1999limits,fogli2003solar,kim2015new,adamson2008measurement,sanchez2018nova}. The first significant result confirming the phenomenon came from the Sudbury Neutrino Observatory (SNO) in 2002\cite{bandyopadhyay2002implications} which provided compelling evidence that all three types of neutrinos (electron, muon, and tau neutrinos) contribute to the solar neutrino flux. Although the discovery by SNO was the turning point but the experiments including SuperKamiokande in Japan and KamLand experiment also played an important role in validating the existence of neutrino oscillations\cite{suzuki2005super,araki2005measurement}. The Super-Kamiokande experiment focused on studying atmospheric neutrinos generated through interactions with cosmic rays while the KamLAND, specifically investigated antineutrinos emitted by nuclear reactors. Recent contributions to the field of neutrino oscillations have been made by a number of cutting-edge experiments some of them are DUNE \cite{escrihuela2017probing} that aims to study neutrino oscillations with unprecedented precision, T2K \cite{miranda2021searching} aims to study the oscillation of muon neutrinos and measure the mixing angle $\theta_{13}$, MINOS/MINOS+, MINOS and its successor, MINOS+, are experiments that use two detectors—one near the neutrino source at Fermilab and the other at a distance in Minnesota—to study neutrino oscillations\cite{adamson2020precision}. MicroBooNE focuses on studying low-energy neutrino interactions and is also involved in neutrino oscillation research \cite{abratenko2023first}. IceCube while primarily designed for studying astrophysical neutrinos but has also made contributions to the field of neutrino oscillations \cite{aartsen2018measurement}. The latest results from NOvA predicts that there exist a strong asymmetry in the rate of $\nu_{e}$ and $\bar{\nu}_e$ appearance thus, inverted hierarchy(IH) at $\delta_{CP}$ = $\pi/2$ more at 3$\sigma$ while normal hierarchy (NH) is highly disfavoured at $\delta_{CP}$ = 3$\pi/2$ at 2$\sigma$\cite{acero2022improved}. DUNE latest results predicts the decoherence parameters $\Gamma_{21}$, $\Gamma_{31}$, $\Gamma_{32}$ at 90 $\%$ C.L. i.e. $\Gamma_{21}$ $\leq$ 5.1 $\times$ $10^{-25}$ GeV, $\Gamma_{32}$ $\leq$ 1.6 $\times$ $10^{-24}$ GeV, $\Gamma_{31}$ $\leq$ 3.0 $\times$ $10^{-25}$ GeV \cite{gomes2019quantum}. The mixing angles as obtained from Superkamiokande experiment $\theta_{12}$, $\theta_{13}$, $\theta_{13}$ lies within [31.27$^\circ$-35.86$^\circ$], [40.1$^\circ$-51.7$^\circ$] and [8.20$^\circ$-8.93$^\circ$] respectively for $3\sigma$ range and that too for normal ordering while for inverted ordering they lies within the range [35.27$^\circ$-35.87$^\circ$], [40.3$^\circ$-51.8$^\circ$] and [8.24$^\circ$-8.96$^\circ$]\cite{esteban2020fate}.

The outcomes of these experiments further helped in confirming and establishing neutrino oscillations as the fundamental phenomena of particle physics. Thus, continuous synergy between the theoretical and experimental efforts has helped us not only to unravel the various properties of neutrinos but also have helped us to understand their role to describe the fundamental laws of nature.

\subsection{The Sterile Neutrinos}
Sterile neutrinos, in theory, are hypothetical neutrinos that do not interact via weak nuclear force, unlike active neutrinos that participate with weak interactions. Sterile neutrinos are proposed to account for several experimental observations that cannot be adequately explained by the three known flavors of active neutrinos (electron, muon and tau neutrinos) within the context of the SM of particle physics. The discrepancy in the results obtained by LSND and MiniBooNE experiments on neutrino oscillations  
provides a compelling motivation for considering the presence of sterile neutrinos \cite{kopp2011there}. The measurement of the hidden decay width of the Z boson at LEP implies that the fourth neutrino flavor is unable to interact via SM weak interactions, thus necessitating it to be a sterile neutrino (3+1). Following the same principle, the model can be expanded by introducing multiple additional neutrino flavor eigenstates. These scenarios fall under the category called as 3+n models. 
If sterile neutrinos are introduced, an extra term to the Lagrangian of SM is to be added and termed as the ``sterile neutrino mass-term'' which can be added as : 
\begin{equation}
     \mathcal{L}= \mathcal{L}_{\text{SM}} + \mathcal{L}_{\text{neutrino}} + \mathcal{L}_{\text{mass}},
\end{equation}
where, $ \mathcal{L}_{\text{SM}}$ is the Lagrangian for SM, $\mathcal{L}_{\text{neutrino}}$ represents the interaction of active neutrinos and $\mathcal{L}_{mass}$ is the sterile neutrino mass term. The said term is written as:
\begin{equation}
    \mathcal{L}_{\text{mass}} = -m_D \cdot N \cdot N^c + \text{h.c.},
\end{equation}
where, $m_{D}$ represents the Dirac matrix, N is the sterile neutrino field, $N^C$ is the charge conjugate field and h.c. denotes the hermitian conjugate term.
The exploration of sterile neutrinos begins in the late 1960s and continued throughout. In the early 1983, Pontecorvo independently proposed the existence of sterile neutrinos which do not engage in weak interactions\cite{pontecorvo1983pages}. In 2000, Pierre Ramond formulated the theory of supersymmetry, a comprehensive framework in particle physics that postulates the presence of extra types of neutrinos, including sterile neutrinos\cite{ramond2000neutrinos}. In 2002, Boris Kayser, Victor Barger and Richard Phillips put forward a framework for three-neutrino mixing that incorporated sterile neutrinos, aiming to provide an explanation for the observed phenomena of neutrino oscillations\cite{de2002can}. In 2004, John Bahcall, Masataka Fukugita and Philip Steinberg conducted a study on the cosmological implications of sterile neutrinos, proposing that these elusive particles could potentially contribute to the enigmatic dark matter that pervades the universe\cite{bahcall2004solar}. In 2006, a group of researchers including Alexei Smirnov, Roberto Peccei and Keith Olive delved into the examination of sterile neutrinos influence on neutrino oscillations, actively exploring diverse experimental signatures associated with their presence \cite{smirnov2006sterile}. 
In 2009, Shmuel Nussinov  investigated how sterile neutrinos would impact neutrino mass models and examined the resulting implications for neutrino oscillations\cite{nussinov2009some}. In 2012, it was Francesco Vissani who investigated the phenomenology and implications of sterile neutrinos not only on neutrino oscillations but also on neutrinoless double beta decay\cite{mitra2012neutrinoless}.

In 2015, the concept of ``neutrinogenesis'' was introduced which proposed that sterile neutrinos might have played a pivotal role in the generation of matter-antimatter asymmetry during the early stages of the universe\cite{heeck2015lepton}. Researchers introduced the ``scotogenic model'' in 2016 as a theoretical framework aimed at providing a comprehensive explanation for both neutrino oscillations and the genesis of dark matter\cite{borah2016light}. The model expanded upon the SM by incorporating sterile neutrinos and introducing a new scalar field.  In 2017, Professor Joseph Lykken and his team put forth the ``neutrino portal dark matter'' model, which put forward the idea that sterile neutrinos could potentially account for the elusive dark matter in the universe\cite{battaglieri2017us,escudero2017sterile}. This model examined the potential for mixing between sterile neutrinos and active neutrinos, investigating its implications for the search for dark matter. In 2018, scientists put forward the ``inverse seesaw mechanism'' as a theoretical framework in response to the challenge posed by the tiny size of neutrino masses and the potential existence of sterile neutrinos\cite{das2018probing}. This mechanism expanded upon the conventional seesaw mechanism and introduced supplementary symmetries, offering a fresh perspective on the origin of neutrino masses. By incorporating these elements, the inverse seesaw mechanism provided a novel description for the observed lightness of neutrinos.
In 2020, theoretical research delved into the prospect of detecting sterile neutrinos based on their gravitational interactions which could influence the dynamics of galaxy clusters, thereby imprinting discernible effects on the overall cosmic structure on grand scale. The researchers extensively examined potential observational signatures that could serve as valuable indicators in the quest for identifying and studying sterile neutrinos\cite{boser2020status}. During the same year, a theoretical study focused on the involvement of sterile neutrinos in the creation of the baryon asymmetry in the universe, utilizing a mechanism known as ``leptogenesis''\cite{garbrecht2020relativistic} and presented a model that incorporated both active and sterile neutrinos, examined the specific conditions that could lead to a substantial asymmetry between matter and antimatter. The study shed light on the potential role of sterile neutrinos in explaining the observed imbalance in the universe's baryon content.

On experimental front, several experiments have been conducted in order to detect sterile neutrinos and the primary objective is to investigate various aspects related to sterile neutrinos, such as their participation in neutrino oscillations, potential connections to dark matter and their role in the matter-antimatter asymmetry of the universe. The LSND (Liquid Scintillator Neutrino Detector) experiment observed an excess of electron antineutrinos, MiniBooNE (Mini Booster Neutrino Experiment) reported intriguing excess electron-like events, the IceCube Neutrino Observatory investigated high-energy neutrinos and found anomalies that could be explained by sterile neutrinos\cite{palomares2005explaining,maltoni2007sterile,aartsen2016searches}. The Daya Bay experiment aimed to measure neutrino oscillations and placed constraints on the existence of sterile neutrinos whereas  SAGE (Soviet-American Gallium Experiment) provided important constraints on sterile neutrinos while primarily focusing on solar neutrinos\cite{an2014search}. Additionally, the MINOS (Main Injector Neutrino Oscillation Search) experiment performed precise measurements of neutrino oscillations, providing valuable insights and placing constraints on the potential presence of sterile neutrinos\cite{adamson2016search}. Results from Daya Bay and MINOS+ collaborations, combined with Bugey-3 exclusion data, yield improved constraints on the \(\theta_{\mu e}\) mixing angle, setting the most constraining limits to date over a broad range of \(\Delta m^2_{41}\), excluding the LSND and MiniBooNE observations at 90\% C.L. for \(\Delta m^2_{41} < 13 \, \text{eV}^2\) and 99\% C.L. for \(\Delta m^2_{41} < 1.6 \, \text{eV}^2\), refining our understanding of neutrino oscillations and sterile neutrino observations\cite{adamson2020improved}.

By investigating the behavior of sterile neutrinos, these experiments offer vital insights into fundamental questions regarding neutrino properties i.e. the existence of dark matter and an extension of the SM. They push the boundaries of our understanding of fundamental particles and their interactions, opening avenues for significant advancements in the fields of particle physics and cosmology. These endeavors contribute to expand our knowledge and shape the future of scientific exploration in these crucial areas of research. 

\subsection{Dark Matter}

Dark matter, is an enigmatic substance that lacks interactions with light or other forms of electromagnetic radiation.

However, the subsistence of dark matter is deduced through observations of how dark matter gravitationally affects visible matter and the overall cosmic structure. It constitutes approximately 85\% of the total matter content in the universe and is undetectable through conventional means.

 An extra term is to be added in the Lagrangian of SM that could account for dark matter by introducing an additional field that represents the dark matter particle. 
\begin{equation}
    \mathcal{L}= \mathcal{L}_{SM} + \mathcal{L}_{DM},
\end{equation}
where, $\mathcal{L}_{SM}$ represents the Lagrangian of SM and $\mathcal{L}_{DM}$ the term for Dark matter field. The $L_{DM}$ is written as: 
\begin{equation}
   \mathcal{L}_{\text{DM}} = 0.5 m_{\text{DM}} \chi^2 - g_{\text{DM}} \chi \bar{\psi}_{\text{DM}} \psi_{\text{DM}},
\end{equation}
where, $m_{\text{DM}}$ represents the mass of the dark matter particle, $\chi$ is the dark matter field and $\psi_{\text{DM}}$ is the field representing the dark matter fermion. The term - $g_{\text{DM}} \chi \bar{\psi}_{\text{DM}}\psi_{\text{DM}}$ represents the interaction between the dark matter field and the dark matter fermion.

The concept of Dark Matter began in the early 1930s, however,
it took decades to get widespread recognition and acceptance within the scientific community. Swiss astronomer Fritz Zwicky was the first one who actually proposed the concept of Dark Matter in 1933\cite{ritchey2011fritz}. While studying Coma Cluster he noticed a wide disparity between the visible mass of galaxies and their observed gravitational effects. He called this ``dunkle Materie'' (dark matter). Till 1970s, no significant work was done on this concept. However, in 1970  American astronomer Vera Rubin performed a series of studies on galaxy rotations\cite{rubin1970rotation}. She observed that stars move at unexpectedly high speeds at the edges of the galaxies, which provided strong evidence of the existence of Dark matter. In 1987, Simon White, an English astrophysicist and Carlos Frenk, a Mexican-British astrophysicist, collaborated on computer simulations and used these simulations to explain the role played by dark matter in the formation of galaxies and galaxy clusters\cite{white1987galaxy}. Their work has made a substantial contribution to our current comprehension of how dark matter shapes the universe on large scales. In 1993, Canadian-American physicist James Peebles developed models of the extensive arrangement of matter in the universe, which includes the presence of dark matter\cite{peebles1993principles}. He established a framework that helped scientists to understand the distribution of dark matter in the cosmos. In the 1994, the theory on cosmic microwave background radiation was proposed by Scott Dodelson, an American theoretical physicist and Marc Kamionkowski, an American cosmologist which provided significant understanding into the composition and distribution of dark matter in the early universe\cite{dodelson1994sterile}. In 1999, three scientists Nima Arkani-Hamed, Savas Dimopoulos and Gia Dvali came up with an interesting idea. They suggested the existence of extra dimensions besides the ones we are familiar with. According to their theory, these extra dimensions could offer an explanation for why gravity appears weaker compared to other forces\cite{arkani1999phenomenology}. This theory introduced new opportunities to comprehend dark matter and its interactions with ordinary matter within the framework of these additional dimensions.

Several models were built by Swedish-American physicist, Max Tegmark in 2004 in order to explain both dark matter and dark energy, which further aims to explain the observed distribution of galaxies \cite{sandvik2004end}. The concept of ``self-interacting dark matter'' was proposed by German astrophysicist Pavel Kroupa in 2006, explained that the particles of Dark Matter interact with each other with forces other than gravity and helped to understand the arrangement of dark matter on small scales\cite{kroupa2008initial}. In 2009, Kathryn Zurek, American theoretical physicist proposed models for the interaction of dark matter with Higgs Boson, thus drawing some possible potential links between dark matter and the SM\cite{kaplan2009asymmetric}.  Gianfranco Bertone, an Italian physicist, focused on understanding what dark matter is composed of, specifically its particles. In his book called ``Particle Dark Matter'' Observations, Models and Searches, which came out in 2010, he explained different ideas about dark matter using theoretical models and observations\cite{bertone2010particle}. In 2011, two American physicists named Dan Hooper and Lisa Goodenough studied information collected by the Fermi Gamma-ray Space Telescope\cite{hooper2011dark}. They noticed that there seemed to be more gamma-ray radiation coming from the middle of our galaxy than Milky Way and suggested that this excess of gamma rays could be a sign of dark matter particles either breaking down or joining together. Their discovery got scientists more interested to find and study dark matter by using indirect methods. In 2020, three physicists named Alexey Boyarsky, Oleg Ruchayskiy and Manuel Tórtola made an interesting suggestion about dark matter. They proposed that dark matter particles could potentially have a tiny bit of electric charge, which would enable them to interact with electromagnetic fields\cite{bondarenko2020direct}. Further, this interaction might be able to explain some unusual observations of X-ray emissions from different objects in space. The idea offered a fresh way of looking at the characteristics and actions of dark matter.
To gain insights into the properties of dark matter and bridge the gaps in our understanding of physics, numerous theories like Cold Dark Matter(CDM) composed of Weakly Interacting Massive Particles (WIMPs) to warm Dark matter (WDM) comprising of light and fast-moving particles have been proposed\cite{de2011warm}. \textit{Axions}, the hypothetical particles addressing strong CP violation gained attention as potential candidates for Dark Matter\cite{duffy2009axions}. Alternatively, theories, like Modified Newtonian Dynamics (MOND) and Self-Interacting Dark Matter (SIDM) have been suggested as possible explanations for the observed phenomena\cite{berezhiani2015theory}. 
In cosmological and astrophysical contexts this theory uses the Boltzmann equation which describes the development of dark matter phase space density function. The Boltzmann equation for SIDM can be written as:
\begin{equation}
    \frac{{\partial f_{\text{DM}}}}{{\partial t}} + \mathbf{v} \cdot \nabla_{\mathbf{x}} f_{\text{DM}} - (\nabla_{\mathbf{x}}\Phi) \cdot \nabla_{\mathbf{p}} f_{\text{DM}} = C[f_{\text{DM}}],
\end{equation}
where, $f_{\text{DM}}(\mathbf{x}, \mathbf{p}, t)$ represents the arrangement of dark matter particles in position (x) and momentum (p) space. $\frac{{\partial f_{\text{DM}}}}{{\partial t}}$ is the time derivative of the distribution function, v is the velocity vector, $\nabla_{\mathbf{x}}$ and $\nabla_{\mathbf{p}}$ are the gradient operators with respect to position and momentum,  $\phi$ is the gravitational potential and $C[f_DM]$ is the collision term that takes into consideration self-interactions terms for dark matter. To incorporate the effect of self-interaction hydrodynamic simulations of SIDM, the fluid equations are modified and uses continuity equation, Euler equation and Poisson equation for the same. The continuity equation represents mass conservation and can be written as:
\begin{equation}
    \frac{{\partial \rho_{\text{DM}}}}{{\partial t}} + \nabla \cdot (\rho_{\text{DM}} \mathbf{v}) = 0,
\end{equation}
where, $\nabla_{\mathbf{x}}$ represents dark matter density and v is the velocity field. The momentum conservation can be described using the Euler equation as:
\begin{equation}
    \frac{{\partial (\rho_{\text{DM}} \mathbf{v})}}{{\partial t}} + \nabla \cdot (\rho_{\text{DM}} \mathbf{v} \otimes \mathbf{v}) = -\nabla P - \rho_{\text{DM}} \nabla \Phi + \mathbf{F}_{\text{inter}},
\end{equation}
where, P is the pressure, $\otimes$ denotes the outer product, $\phi$ is the gravitational potential and $F_{inter}$ represents the additional force due to self-interactions. 

On an astrophysical scale, a number of techniques have been employed to study the dark matter effects. Bending of light by massive objects, Gravitational lensing has provided compelling results for the existence of dark matter on both cosmological and galactic scales\cite{massey2010dark,huterer2015growth}. Furthermore, the measurements made by Planck Satellite via cosmic microwave background radiation play an important role in determining the cosmological parameters, including the amount of dark matter in the universe\cite{di2017can}.

On the experimental side, several experiments have been performed to detect dark matter. The detection of dark matter through direct methods such as XENON1T and XENONnT,  LUX (Large Underground Xenon),  DAMA/LIBRA,  
 CDMS (Cryogenic Dark Matter Search) includes the interplay between particles of dark matter and regular matter whereas indirect methods like Fermi-LAT, HESS (High Energy Stereoscopic System), IceCube and AMS-02 (Alpha Magnetic Spectrometer) involve the detection of left out products after the annihilation of dark matter\cite{aprile2017first, aprile2020projected, szydagis2014detailed, ahmed2009dark,boucenna2015decaying,budrikis2021decade}. The collider experiments performed in LHC CERN aim to produce dark matter through collision products and its investigation for the missing signals\cite{thomas2008mixed}. Recent experiments that have made significant contributions to the existence of dark matter include DES which focuses on dark energy, maps large-scale structures in the universe by studying the gravitational lensing effect and helps in understanding the properties of dark matter, XENONnT and LZ: XENONnT and LZ (LUX-ZEPLIN) aim to detect the dark matter directly through interactions with dark matter, DAMIC that make the use of charged coupled devices in order to search low mass dark particles, CRESST that uses superconducting materials to search for dark matter \cite{chen2021constraints,catena2018compatibility,lee2020dark,petricca2020first}. The latest results as obtained from XENONnT that for spin-independent interactions, the lowest upper limit comes out to be 2.58$\times$ $10^{-47}$ $cm^{2}$ at 28 GeV/$c^2$ with 90 $\%$ C.L. and for masses above 100 GeV/$c^2$ the limit is 6.08$\times$ $10^{-47}$ $cm^{2}$ $\times$ [$M_{DM}$/(100 GeV/$c^2$ )]\cite{aprile2023first}. 
 
 Understanding dark matter is essential to get deeper insights into the processes of structure formation, galaxy evolution and the overall cosmic web. Also, it comprehends the underlying mechanisms that have shaped the universe's history and continue to govern its evolution. Thus, knowing dark matter will complete our picture of the universe and explains the missing mass that shapes its structure.

\subsection{Non-Standard Interactions: (NSIs)}

Non-standard Interactions (NSIs) are additional interactions that extend beyond the weak force predicted in the SM of particle physics. They originate from novel particles or forces and have the potential to impact the behavior of neutrinos in various ways such as NSIs can introduce modifications to neutrino oscillation probabilities, influence the rates at which neutrinos are detected in experiments, and potentially offer explanations for observed deviations from the standard three-neutrino oscillation framework. By considering NSIs, scientists can explore the effects of these additional interactions on neutrino propagation and interactions, providing valuable understanding of fundamental properties of neutrinos and the possibility of physics beyond the SM.

The concept of NSIs started in mid 20th century i.e. in 1977 by Lincoln Wolfenstein in a paper titled ``Neutrino Oscillations in Matter'' which proposed that neutrinos could interact with matter in a way that was not predicted by the SM of particle physics\cite{wolfenstein1978neutrino}. This could lead to observable effects in neutrino oscillation experiments, such as an increase in the neutrino oscillation probability or a change in the neutrino energy spectrum. The concept of split supersymmetry was proposed by  Howard Georgi and Savas Dimopoulos in 1986 where they introduced the new sources of flavor violation that affect NSIs\cite{georgi1986theory,hernandez2001trilinear}. A theory called ``Effective Field theory'' was proposed by Steven Weinberg in 1991 in which he described how NSIs affect neutrino oscillations and provided a powerful tool for analyzing experimental data\cite{xing2004flavor,weinberg2008effective}. P. I. Krastev and Alexander Yu. Smirnov in 1998 examined the possibility how neutrino ocillations gets affected in the presecnce of neutrino oscillations\cite{joshipura1998rescuing}. The idea of NSI was further developed in the 1999 by a number of physicists, Sidney Coleman, Sheldon Glashow and Hitoshi Murayama and explored the deviation of standard interactions of neutrinos with matter\cite{coleman1999high}.

In 2001,  Pilar Hernández, Francisco López and Javier Bernabéu investigated NSIs in relation to the Minimal Supersymmetric Standard Model (MSSM) and studied the effects of NSIs on processes such as muon decay and proposed methods for searching for NSIs in future experiments\cite{gavela2009minimal}.  
In 2007, Carlo Giunti and Chung W. Kim provided an extensive overview of neutrino physics including discussions on NSIs in their book titled ``Fundamentals of Neutrino Physics and Astrophysics'' \cite{giunti2007fundamentals}.

In 2008, Stephen Parke and Toshihiko Takaki studied the effect of NSIs on neutrino oscillations, particularly with regard to the future Long-Baseline Neutrino Experiment (LBNE) in the United States \cite{nunokawa2008cp}. A parameterization was proposed by Enrique Fernandez-Martinez, Joachim Kopp and Michele Maltoni in 2009 to describe these interactions in a model-independent manner\cite{antusch2009non}. Joachim Kopp, Pedro Machado and Michele Maltoni in 2012 focused mainly on the potential impact of NSIs on long-baseline neutrino experiments, particularly MINOS and T2K experiments \cite{gonzalez2013determination}. An attempt was made by Andrea Donini, Pilar Hernández, Juan Carlos Romão and Martin Hirsch in 2013 to show a connection between non-standard interactions and the presence of sterile neutrinos\cite{dorman1981cosmic}. The phenomenological implication of NSIs on dark matter was considered by Céline Bœhm, Dorota Jurciukonis and Neill Raper in 2015, considering the effects of NSIs on dark matter detection experiments and its detection through indirect means in astrophysical observations\cite{gluscevic2019cosmological,malik2015interplay}. Consequences of NSIs on neutrino-electron scattering experiments and proposed experimental strategies to probe the presence of NSIs were discussed by Andrés Ayala, Alvaro Hernández-Cabezudo and Joachim Kopp in 2020\cite{fischer2020explaining}.

Thus, in order to explain the general formalism of NSIs, the Lagrangian for both charged and neutral current NSIs describing these interactions is given as:
\begin{equation}
   \mathcal{L}_{NSIs} = \epsilon_{f_{\alpha\beta}} (\nu_\alpha\gamma^\mu P_L \nu_\beta) (f \gamma_\mu P_f),
\end{equation}
where, $\nu_{\alpha}$ and $\nu_{\beta}$ are the flavor eigenstates, f is the fermion species in matter and $P_{L}$ is the left-handed projection operator. $\epsilon_{f_{\alpha\beta}}$  tells the strength of the NSIs for each fermion species and neutrino flavor combination.
The impact of NSIs on neutrino oscillations can be seen only by modifying probabilities of oscillations among three flavors. The modified probabilities due to NSIs can be expressed as:
\begin{equation}
    P(\nu_\alpha \to \nu_\beta) = |U_{\alpha i}|^2 |U_{\beta i}|^2 + 2 \operatorname{Re}[\epsilon_{f_{\alpha\beta}} U_{\alpha i} U_{\beta i}^* U_{fi} U_{fi}^*] + \mathcal{O}(\epsilon^2),
\end{equation}
where, $U_{\alpha i}$ and $U_{\beta i}$ represents the components of the neutrino mixing matrix. The modified probabilities depend upon the mixing angles, mass squared differences and the NSIs parameter $f_{\alpha\beta}$. And, if neutrinos oscillate through matter, the scattering cross sections between  neutrino-electron or neutrino-nucleon scattering is given as:
\begin{equation}
    \sigma(\nu_\alpha e \rightarrow \nu_\beta e) = G_F^2 |\epsilon_{e_{\alpha\beta}}|^2,
\end{equation}
\begin{equation}
    \sigma(\nu_\alpha N \rightarrow \nu_\beta N) = G_F^2 |\epsilon_{N_{\alpha\beta}}|^2,
\end{equation}
where, $G_{F}$ is the fermi constant $\epsilon_{e\alpha\beta}$ and $\epsilon_{N\alpha\beta}$ are the NSI parameters for electron and nucleon interactions, respectively. On the other hand, the oscillation probabilities also get affected due to NSIs when neutrinos propagate through matter can be expressed using the equation governing the evolution of neutrino flavor states.  Both the standard matter-induced oscillations and the additional effects of NSIs are included under the evolution equation given as
\begin{equation}
    P(\nu_\alpha \to \nu_\beta) = \left| \sum_{k} U_{\alpha k} e^{-i\frac{m_k^2L}{2E}} U_{\beta k}^* \right|^2 + \sum_{j\neq k} 2 \operatorname{Re} \left[ \left( \sum_{f} \epsilon_{\alpha f} U_{\alpha f} U_{\beta f}^* \right) e^{-i\frac{\Delta m_{jk}^2L}{2E}} \right].
\end{equation}
Experimental constraints on NSIs can be obtained from several sources, including neutrino oscillation experiments, neutrino scattering experiments and astrophysical observations.
Constraints on NSI have been established through the findings of multiple experiments. In the past, experiments such as the LSND and MiniBooNE have provided hints of potential NSIs by observing an abundance of electron neutrinos compared to muon neutrino beam\cite{liao2016impact,akhmedov2011new}. Notably, the MINOS and NOvA experiments have investigated neutrino oscillations over extended distances to place limitations on NSIs\cite{sousa2015first,coelho2017nonmaximal}. Through their accurate measurements of neutrino oscillation parameters, the T2K and Daya Bay experiments have made significant contributions to the constraints on NSIs\cite{girardi2014daya}. Recent experiments have made significant contributions to the field of NSIs by advancing our understanding of the phenomena through their innovative approaches and insightful findings are SBND that explores the neutrino anomalies\cite{machado2019short}, COHERENT which focuses on coherent elastic neutrino-nucleus scattering (CEvNS) and is sensitive to NSIs\cite{giunti2020general,agarwalla2012constraining}. From NOvA experiment it is found that neutrino mixing in atmospheric sector deviates from maximum \(
\sin^2 2\theta_{23} = 0.5
\)
by \(2.6\sigma\)
whereas the results from COHERENT experiment that the cases where the NSIs are due to up and down quarks only excludes the LMA- Dark at 5.6$\sigma$ and 7.2$\sigma$ respectively\cite{giunti2020general}.

In conclusion, by analyzing the energy spectrum, flavor composition and scattering patterns of neutrinos, experiments aim to uncover new physics phenomena and gain insights into neutrino properties. This research aims to detect the presence of NSIs, which would indicate the existence of physics beyond the SM and expand our understanding of fundamental particles. Through collaborative endeavors in experimentation and theoretical calculations, scientists can constrain NSIs and gain insights into the nature of neutrinos and their interactions with matter. Their findings have the potential to revolutionize our understanding of particle physics and cosmology, opening up new frontiers in our exploration of the fundamental fabric of reality.

\subsection{Neutrino mass in Extra Dimensions}

The concept of extra dimensions in theoretical physics suggests that the particles outlined in the SM have the ability to move and exist in spaces beyond the usual three dimensions. In these scenarios, the presence of additional dimensions can give rise to novel mechanisms responsible for generating the masses of neutrinos and determining their mixing patterns. The idea of extra dimensions can bring about new particles or interactions that contribute to the creation of neutrino masses. These additional elements can explain why neutrino masses are so small or why  neutrinos mix together is peculiar. The concept emerged in the early 20th century but gained importance in the late \textit{20th} century after the discovery of neutrino oscillations. In 1994, Ignatios Antoniadis and Kostas Benakli investigated the potential for creating minute Majorana masses for neutrinos in theories with extra dimensions\cite{antoniadis1999soft,antoniadis1994exact}. They proposed that the minuteness of neutrino masses could be achieved by reducing the strength of certain interactions through the size of the extra dimension. In 1995, Georgi Dvali and Alexander Yu. Smirnov proposed a mechanism for generating small neutrino masses in theories with large extra dimensions\cite {arkani2002large,joshipura1998rescuing}. They explored scenarios where sterile neutrinos move through the extra dimensions and interact with particles of the SM, leading to the observed pattern of neutrino masses. 

In 1998, Nima Arkani-Hamed, Savas Dimopoulos and Gia Dvali came up with the ADD model, which introduced the idea that there might be dimensions larger than the ones we usually see\cite{antoniadis1998new}. They wanted to solve the hierarchy problem of neutrinos, but their model had implications for how neutrinos, tiny particles with no electric charge, get their mass. Finally, in 1999, Lisa Randall and Raman Sundrum introduced the idea of a warped extra dimension in their model, which aimed to solve the hierarchy problem\cite{randall1999large}. Their work had implications for understanding the mass of neutrinos within the given context of extra dimensions, opening up new possibilities for further investigation. In 2003, Ignatios Antoniadis, Kostas Benakli and Mikael Laletin explored the idea of having small masses for neutrinos by thinking about gravity at a very high energy scale and having extra big dimensions\cite{antoniadis2003physics}. They suggested situations where specific categories of neutrinos traverse through these additional dimensions, resulting in the emergence of their small masses. 

Zhi-Zhong Xing and Shun Zhou looked at how neutrino masses and the way they mix together could happen in the context of these extra dimensions in 2004\cite{xing2004isomeric}. They studied different ways the neutrinos can move and how they interact with the Higgs field, which give particles their mass, in these extra dimensions. In 2007, Alfredo Aranda, Julio C. Montero and Carlos Rojas examined the implications for neutrino masses and mixing that would arise from the presence of extra dimensions.\cite{aranda2007neutrino}. They made a model with five dimensions where neutrinos move through them in extra dimensions which could explain the masses and the mixing patterns so observed. In 2008, Ernest Ma proposed a five-dimensional model that introduced right-handed neutrinos in the bulk, explaining the smallness of neutrino masses through the suppression of their wave function overlaps with the Higgs field\cite{ma2008z3}. Also in 2008, Csaba Csáki, Christophe Grojean and Joel Wacker investigated warped extra dimensions and proposed scenarios where the observed neutrino mass hierarchy arises from the interplay between bulk mass parameters and the localization of fermion fields\cite{csaki2008model}. Finally, in 2011 Kei Yagyu and Kento Yoshida explored the implications of the Randall-Sundrum model, proposing a scenario where mixing angles and mass differences of neutrinos can be naturally generated through the interplay of bulk mass parameters, brane-localized Yukawa couplings and brane-localized Majorana masses\cite{kanemura2011non}.

Although the explanations for mass generation in extra dimensions depend upon the specific model considered, the general approach and overview for the same is the ``brane-bulk'' scenario where the SM particles are restricted to a four-dimensional brane, while gravity can extend through the higher-dimensional bulk. Here, the neutrinos are assumed to have left-handed chiralities that propagate on the brane, while the right-handed neutrinos propagate in the extra dimensions. The effective four-dimensional Lagrangian in the brane-bulk scenario describing the neutrino mass terms  for the higher-dimensional theory is described as
\begin{equation}
    \mathcal{L}_{\text{{mass}}} = -\frac{1}{2} \left(\overline{\nu}_{L} M_{\nu} \nu_{R} + \text{{h.c.}}\right),
\end{equation}
where, the mass matrix $M_{\nu}$ depends upon the symmetries and interactions in the higher-dimensional theory. Calculation of $M_{\nu}$ depends upon the specific model, the number and nature of the extra dimensions and the symmetries and interactions present in the higher-dimensional theory but one of the approaches to calculate the effective Lagrangian and neutrino mass terms is the Kaluza-Klein expansions. The Kaluza-Klein expresses the right-handed neutrino field in extra dimensions as:
\begin{equation}
    \nu_R(x, y) = \sum_{n=0}^\infty \frac{1}{\sqrt{2\pi R}} \nu_R^{(n)}(x) \chi_n(y),
\end{equation}
where, $\nu_{R}$ is the five-dimensional right-handed neutrino field, $\nu_R^{(n)}(x)$ are the four-dimensional Kaluza-Klein modes, $\chi_n(y)$  represents the profiles of the right-handed neutrino in the extra dimensions and R is the radius of the compactified extra dimensions. On the other hand, the mass matrix $M_{\nu}$ gets contributions from both the zero and the Kaluza-Klein modes. The overlap of the left-handed and zero-mode right-handed neutrinos leads to the zero mode contribution and $M_{\nu}$ acquires the form
\begin{equation}
    (M_{\nu})_{ij} = \frac{1}{\sqrt{2\pi R}} \int dy \, \chi_0(y) \, \left[\lambda_{ij}^{(0)} \, \phi(x) + \sum_{n=1}^{\infty} \lambda_{ij}^{(n)} \, \phi^{(n)}(x) \right],
\end{equation}
$\lambda_{ij}^{(0)}$, $\lambda_{ij}^{(n)}$  are the Yukawa couplings associated with zero and the Kaluza-Klein modes and $\phi(x)$ is the Higgs field and $\phi(x)$ and $\phi^{(n)}(x)$ are the Higgs fields in Kaluza-Klein mode.

Behind the theoretical framework on neutrino mass in extra dimensions, a number of experiments have been performed and several are in running phase.
The experiments like LSND, KamLAND, IceCube, NEOS, NOvA play an important role in probing the existence of extra dimensions and investigating their potential impact on neutrino mass\cite{davoudiasl2002constraints,machado2011testing,esmaili2014probing,ko2017sterile,backhouse2015results}.
These experiments have provided the most constructive bound from the data obtained from atmospheric data in the hierarchical mass scheme, constrains the largest extra dimension to have a radius R $< 0.82 {\mu}m$. When the lightest neutrino mass approaches zero, a stronger constraint on the size of the extra dimension emerges from the combined data of CHOOZ, KamLAND and MINOS. For both normal and inverted mass hierarchies, this constraint is approximately $1.0\,(\pm 0.6)\,\mu\text{m}$ at a 99 $\%$ C.L. In the case where the lightest neutrino mass is larger, say 0.2 eV, for instance, the constraint is around $0.1\,\mu\text{m}$\cite{machado2011testing}.

Till now, none of the experiment has provided a positive definitive signal specifically indicating the existence of neutrino mass in extra dimensions. A number of experiments like DUNE, JUNO and Hyper-Kamiokande have been planned for the future, to provide further insights into the possible connection between neutrino mass and extra dimensions\cite{roy2023capability,basto2022short,bian2022hyper}.
Studying neutrino mass in extra dimensions will not only provide information about the nature of neutrinos but will also offer alternate approaches to resolve enigmas like dark matter, the matter-antimatter imbalance in the universe and the generation of neutrino masses. All these BSM theories which explores the extensions of SM, act as intermediate steps to achieve the ultimate goal ``The Theory of everything''.

\section{Grand Unified Theories (GUT)}

The main motivation behind the BSM theories is the unification of the fundamental forces of nature: the electromagnetic force, the weak nuclear force and the strong nuclear forces. In SM, these forces are described by different theories, however at extremely high energies \(\sim \) $10^{14}$ GeV they become unified and can be described by a single theory called Grand Unified Theory(GUT).
It is a mathematical concept that relies on gauge symmetry. The symmetry is broken during the cooling phase following the Big Bang, causing the unified force to split into the distinct forces we observe today. 

The concept of GUT began in the late 1970s and in early 1971  physicists Sheldon Glashow, Howard Georgi and Richard Slansky came up with an idea called the SU(5) model\cite{robinson1979diamond}. This model aimed to bring together the electromagnetic and the weak nuclear force. They proposed a unified gauge theory, which means they tried to explain how these two forces are connected and can be understood as part of a single force. Howard Georgi, Sheldon Glashow, Makoto Machacek and Toichiro Matsuura in 1974, worked together to expand SU(5) model\cite{georgi1974unity}. They delved deeper into its details and studied how it affects the interactions and predictions of particles. In 1974, SU(6) model was proposed by Benjamin W. Lee and Quang Ho-Kim which not only extended the SU(5) symmetry but also introduced some additional gauge bosons\cite{lee19773}. The models were further extended to expand on the ideas of Georgi and Glashow. In 1974-1975, physicists John Ellis, Mary Gaillard and Dimitri Nanopoulos introduced the concept of supersymmetry into GUTs\cite{ellis1983primordial} which suggests a connection between two types of particles, bosons and fermions and it offers potential solutions to certain issues within the SM. Nathan Seiberg and Edward Witten in 1985 introduced the concept of Seiberg-Witten theory, which helped in understanding supersymmetric gauge theories and their connections to GUTs\cite{affleck1985dynamical}. This research laid the foundation for progress in string theory and the concept of gauge/gravity duality, opening up new avenues for further exploration and understanding in these fields\cite{dienes1996realizing}. In 1986, Howard Haber and Gordon Kane made significant progress in the development of GUTs by introducing the Minimal Supersymmetric SM (MSSM)\cite{quiros1986non}. The MSSM took GUT theories a step further by incorporating the concept of supersymmetry. This framework aimed to unify all the fundamental forces and particles known in nature, providing a comprehensive and cohesive understanding of the universe. Joel Primack, George Blumenthal and Martin Rees in 1987 suggested that GUTs might offer an explanation for the mysterious existence of dark matter\cite{rees1987dark}. Their research focused on investigating the potential link between the unification of fundamental forces within GUTs and the presence of non-baryonic dark matter particles, which do not consist of ordinary matter like protons and neutrons.

  In 1999, Z Kakushadze contributed to the field of string phenomenology, focused on exploring the potential for constructing realistic models of particle physics within the framework of GUTs\cite{kakushadze1998review}.
 The Randall-Sundrum model was proposed by Lisa Randall and Raman Sundrum in 1999, within the context of extra dimensions and addressed the hierarchy problem in particle physics\cite{randall1999out}.  In 2002, Daniel Friedan, Emil Martinec and Stephen Shenker made significant contributions by introducing the concept of Duality Symmetry\cite{kaloper2002signatures}. In 2006, Hirosi Ooguri, Cumrun Vafa and Erik Verlinde introduced the idea of Gauge-Gravity Duality, commonly known as the AdS/CFT correspondence which established a profound connection between specific gravitational theories and gauge theories\cite{dijkgraaf2006m}. In 2007, Nima Arkani-Hamed and his collaborators introduced the concept of ``emergence'' in particle physics which explored the concept that the fundamental laws of nature can arise from simple underlying principles\cite{arkani2007string}. Their work hold implications for the comprehension of GUT and the unification of forces in the universe. Yasunori Nomura in 2008 put forward fresh perspectives and innovative approaches to the development of theories involving hidden sectors and their implications for GUT\cite{nomura2008flavorful}. His work suggested novel ways to explore the unification of forces. 
 
 In 2009, Cumrun Vafa comprehended F-theory, a branch of string theory\cite{beasley2009guts} which incorporate  extra dimensions, presenting exciting prospects for the unification of fundamental forces in the universe. During the same year, Maxim Perelstein also put forward innovative approaches to detect potential signatures of GUT at the Large Hadron Collider (LHC)\cite{csaki2009weakly}. His proposals offered valuable insights into experimental tests and the exploration of GUT predictions. Perelstein's work introduced novel methods for investigating the presence of GUT and their manifestations in high-energy particle collisions at the LHC\cite{farina2014higgs}. Connections between GUT and dark matter were being investigated which explored the possibility that dark matter particles could be related to the breaking of GUT symmetries by Howard Baer, Vernon Barger and other researchers in 2012\cite{baer2012natural}. While, in 2013, Riccardo Torre studied the relationship between GUT and the Higgs boson\cite{franceschini2013rpv}. They looked into how GUT symmetries contribute to the creation of particle masses and the functioning of the Higgs mechanism. In 2014, Yang-Hui He and his colleagues examined the mathematical patterns of GUT\cite{he2014heterotic}. They specifically looked at how exceptional groups play a crucial role in creating a unified structure for the fundamental forces.  In 2015, Timothy T. Yanagida put forward an idea called ``leptogenesis'', helps to explain what accounts for the predominance of matter over antimatter in the universe, by connecting it to GUT\cite{mcdonald2015radiatively}.
    In 2020, Y Lozano, Carlos M. Nuñez and their team of researchers introduced a fresh approach to explain the breakdown of GUT\cite{lozano2020m}. They suggested using higher-dimensional geometric structures called branes to achieve this. Their findings offer exciting possibilities for comprehending GUT within the framework of string theory and extra dimensions.

Several important experiments have greatly helped us to understand GUTs. The Super-Kamiokande experiment aimed at detecting proton decay as predicted by GUTs, stringent constraints were established on the potential lifespan of protons\cite{takhistov2016review}. The  LHC at CERN has been crucial in smashing particles together at very high energies. This gives us data to test GUT predictions and improve our theories. Experiments with neutrinos like SNO have taught us about neutrino properties such as their mass and how they change types. This is important for GUTs too. Precise measurements of the electroweak force at the LEP and Tevatron accelerators have helped us to understand how forces unify at higher energies\cite{fustin2003motivation, croon2019gut,ellis1990precision,baer2002reach}. All these experiments have contributed a lot to our knowledge of GUTs and our understanding of how the basic forces work in the universe.

  In conclusion, GUT stands as a captivating framework that aims to integrate the fundamental forces of nature and provide insights into the early universe and its evolution. It can provide information about the fundamental processes that shaped our universe from its earliest moments and can provide explanations for various phenomena such as the 
prevalence of matter over antimatter, the origin of particle masses and the formation of cosmic structures. At last, studying GUTs offers a chance to discover and advance mathematical techniques. This exploration not only enriches our understanding of physics but also contributes to the broader realm of mathematics.

\section{Conclusion}

 This article examined various crucial elements of modern particle physics, its basic principles and advancements beyond the Standard Model. Gauge invariance, as demonstrated by Yang-Mills theory, plays a significant role in our understanding of the fundamental forces in nature. The model given by Glashow-Weinberg-Salam (GWS) has successfully united the electromagnetic and weak forces, offering a comprehensive framework to explain interactions between particles. Further, the phenomenon of spontaneous symmetry breaking played a crucial role in elucidating the origin of particle masses.  
The observation of neutral current interaction as well as discovery of $W^\pm$
, $Z^{0}$ at CERN were crucial experiments which gave impetus to SM, explaining the over increasing particle data. Despite of its startling success in its credit, SM falls short of being called as the fundamental theory. The observation of neutrino oscillations provides unambigous signal to go BSM. Vast amount of efforts have been put in to unravel the physics that goes beyond the Standard Model in the hope of arriving at the so called ``Theory of Everything''. This journey showcase the persistent pursuit of unraveling the mysteries of particle physics, with the aim of grasping the fundamental principles that govern the universe. These advancements will lead us closer to fully understand the basic nature of the universe and the particles that make it up.
\section{Acknowledgement}
Lisha and Neelu Mahajan would like to express our sincere gratitude to the Principal, GGDSD College, Chandigarh for providing facilities to work. Suyash Mehta would like to extend sincere thanks to Director, Indian Intitute of Space Science and Technology Valiamala Thiruvananthapuram, Kerala for granting the opportunity to utilize the facilities necessary for work.
\bibliographystyle{unsrt}
\bibliography{sample.bib}
\end{document}